\documentclass[final,11pt]{article}

\usepackage{multirow}
\usepackage[nottoc]{tocbibind}
\usepackage{geometry}
\usepackage{subcaption}
\usepackage[affil-it, auth-sc]{authblk}
\usepackage[english]{babel}
\usepackage[utf8]{inputenc}
\usepackage{cite}
\usepackage{graphicx}
\usepackage{amssymb}
\usepackage{amsthm}
\usepackage{amsmath}
\usepackage{amsfonts}
\usepackage{verbatim}
\usepackage{lineno}
\usepackage{todonotes}
\presetkeys{todonotes}{inline, color=blue!30, size=\small}{}

\usepackage{color, soul}
\definecolor{darkred}{rgb}{0.9, 0.0, 0.0}
\definecolor{darkgreen}{rgb}{0.0, 0.5, 0.0}

\usepackage[colorlinks,bookmarks,linkcolor=darkred, citecolor=darkgreen]{hyperref}
\usepackage{eso-pic}
\usepackage[sort&compress,numbers]{natbib}
\usepackage{doi}
\usepackage{paralist,epsfig}
\usepackage{relsize}
\usepackage{nicefrac,esint}
\usepackage{float}
\usepackage{cleveref}

\begin{document}

\AddToShipoutPictureFG*{
 \AtPageUpperLeft{\put(-60,-60){\makebox[\paperwidth][r]{CETUP2025-011}}}}

\title{Reducing Hadronic Uncertainty in Low-Energy Neutral-Current Processes}

\author{Oleksandr Tomalak \thanks{tomalak@itp.ac.cn}}
\affil{Institute of Theoretical Physics, Chinese Academy of Sciences, Beijing \& 100190, P. R. China}

\date{\today}

\maketitle

\begin{abstract}
We analyze the hadronic uncertainty from light-quark loops coupled to (anti)neutrino in low-energy neutral-current (anti)neutrino scattering, estimated at the $3$-$4$ permille level. This uncertainty arises from limited knowledge of the charge-isospin correlation function of quark currents. We study the charge-charge and charge-isospin correlators within $\mathrm{SU}(2)$ and $\mathrm{SU}(3)$ chiral perturbation theory (ChPT). In $\mathrm{SU}(2)$ ChPT, the two correlators are identical to all orders in the chiral and electromagnetic expansions. We further perform a leading-order $\mathrm{SU}(3)$ ChPT calculation and discuss the relevant counterterms. Our findings reduce the hadronic uncertainty in neutral-current processes such as (anti)neutrino-electron and coherent elastic (anti)neutrino-nucleus scattering by a factor $\sim 35$.
\end{abstract}

\begin{figure}[htb]
	\centering
	\includegraphics[width=0.49\textwidth]{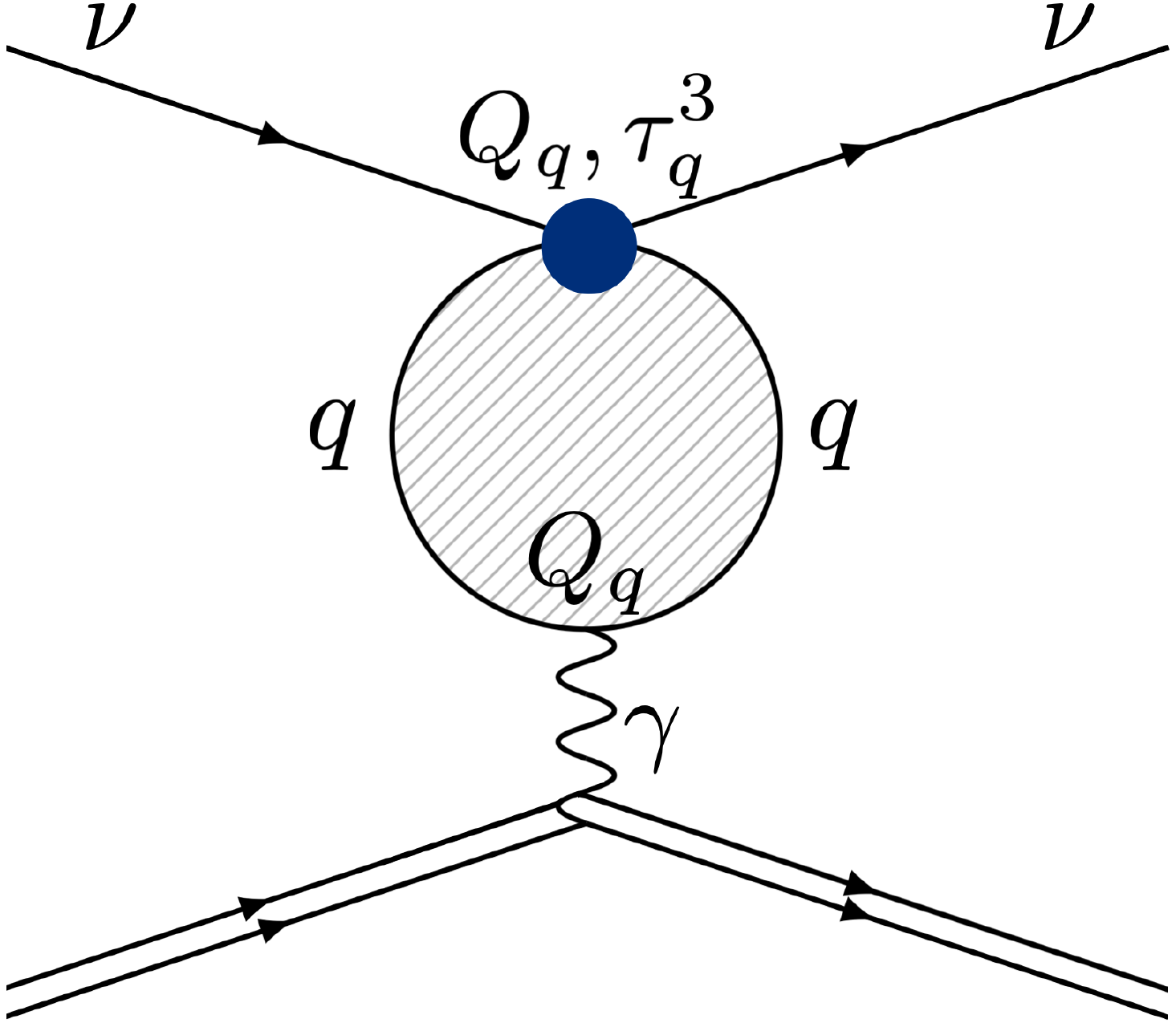}
	\caption{Non-perturbative flavor-independent closed fermion loop contribution to neutral-current (anti)neutrino scattering that involves light quarks is illustrated.} \label{fig:diagram}
\end{figure}

Modern experimental tests of the Standard Model (SM) have reached percent and subpercent precision in both leptonic and hadronic sectors~\cite{Antognini:2013txn,Pohl:2010zza,NA62:2012lny,A1:2010nsl,A1:2013fsc,MINERvA:2015nqi,Super-Kamiokande:2016yck,CMS:2018ktx,Markisch:2018ndu,MINERvA:2019hhc,UCNt:2021pcg,DayaBay:2022orm,Muong-2:2025xyk}. This level of accuracy requires precise theoretical control over higher-order corrections in perturbation theory. Electroweak radiative corrections are typically organized in an expansion in the electromagnetic coupling $\alpha = \frac{e^2}{4 \pi}$, where $e$ is the electric charge, with coefficients determined by SM parameters. In contrast, strong interaction effects can be treated perturbatively only at high energies, where the strong coupling $\alpha_S$ is small. Electroweak radiative corrections to neutral-current (anti)neutrino scattering were initially studied in Refs.~\cite{Ram:1967zza,Byers:1979af,Green:1980uc,Marciano:1980pb,Aoki:1980ix,Aoki:1981kq,Hioki:1981gi,Sarantakos:1982bp,Bardin:1983yb,Bardin:1983zm,Bardin:1985fg,Passera:2000ug} based on the tree-level results in Refs.~\cite{Weinberg:1967tq,tHooft:1971ucy}. With precise knowledge of SM parameters, the remaining theoretical uncertainties in low-energy (anti)neutrino scattering are dominated by hadronic and nuclear effects. Of particular interest are hadronic contributions that enter through loop diagrams involving light quarks, which couple to neutrinos via short-distance interactions, and to the target through photon exchange, as illustrated in Fig.~\ref{fig:diagram}. Earlier estimates~\cite{Marciano:1980pb,Sarantakos:1982bp,Marciano:1983ss,Hollik:1988ii,Denner:1991kt,Ferroglia:2003wa,Kumar:2013yoa,Erler:1998sy,Erler:2004in,Erler:2017knj,Huang:2024rfb} of such hadronic effects relied on free-quark models with effective massive quarks. This approximation becomes inadequate at low energies where quarks are no longer valid degrees of freedom. Below the quantum chromodynamics (QCD) confinement scale, perturbative expansion in $\alpha_S$ breaks down, and hadronic contributions must be treated non-perturbatively. A field-theoretic formulation avoiding quark mass models was proposed in~\cite{Jegerlehner:1985gq,Degrassi:1989ip,Jegerlehner:2011mw}, employing current-current correlators constructed from light-quark currents. Two such correlators are relevant: the charge-charge and the charge-isospin correlators. While the former can be extracted from experimental data with subpercent precision~\cite{Erler:2017knj,Jegerlehner:2019lxt,ParticleDataGroup:2020ssz}, the latter cannot be directly accessed and is thus less well-known. Yet, it is the charge-isospin correlator that dominates the theoretical uncertainty in numerous low-energy neutral-current (anti)neutrino scattering processes, such as (anti)neutrino-electron and coherent elastic (anti)neutrino-nucleus scattering (CEvNS).

The charge-isospin correlator contributes universally to neutral-current processes involving leptons and quarks. It also plays a central role in interpreting low-energy parity-violating electron scattering experiments~\cite{SLACE158:2005uay,Qweak:2013zxf,PVDIS:2014cmd,Qweak:2018tjf,Becker:2018ggl,MOLLER:2014iki,Chen:2014psa,Souder:2016xcn}, where its uncertainty limits the interpretation of electroweak observables~\cite{Marciano:1983ss,Czarnecki:1995fw,Czarnecki:1998xc,Czarnecki:2000ic,Ferroglia:2003wa,Kumar:2013yoa,Erler:1998sy,Erler:2004in,Erler:2017knj}. Following the methodology developed for evaluating hadronic contributions to the running of the electromagnetic coupling~\cite{Bernecker:2011gh,Ce:2022eix} and the muon anomalous magnetic moment~\cite{Blum:2002ii,Bernecker:2011gh,Chakraborty:2016mwy,Budapest-Marseille-Wuppertal:2017okr,FermilabLattice:2017wgj,RBC:2018dos,Aoyama:2020ynm,ExtendedTwistedMass:2022jpw,Ce:2022kxy,FermilabLatticeHPQCD:2023jof,RBC:2023pvn,Boccaletti:2024guq,Kuberski:2024bcj,RBC:2024fic,Spiegel:2024dec,Djukanovic:2024cmq,Aliberti:2025beg,Erb:2025nxk}, the charge-isospin correlator has been recently evaluated by a simulation of the non-perturbative strong dynamics of quarks and gluons, QCD, on a space-time lattice~\cite{Burger:2015lqa,Francis:2015grz,Ce:2018ziv,Ce:2022eix}.

In addition to a proper effective field theory (EFT) treatment of the (anti)neutrino-electron scattering, Refs.~\cite{Tomalak:2019ibg,Hill:2019xqk} formulated light-quark contributions to neutral-current (anti)neutrino scattering in terms of well-defined non-perturbative hadronic objects of Refs.~\cite{Jegerlehner:1985gq,Jegerlehner:2011mw}, corrected the normalization by a factor of 2 in earlier definitions of the charge-isospin correlation function, and quantified the uncertainty of the ``pure-leptonic" (anti)neutrino-electron scattering process. Subsequently, these methods were extended to the recently-discovered coherent elastic (anti)neutrino-nucleus scattering~\cite{Stodolsky:1966zz,Freedman:1973yd,Kopeliovich:1974mv,Sehgal:1985iu,Botella:1986wy,COHERENT:2017ipa,Tayloe:2017edz,COHERENT:2022nrm,COHERENT:2020iec,COHERENT:2024axu,RED-100:2024izi,Ackermann:2025obx} in Ref.~\cite{Tomalak:2020zfh}, with detailed error analysis. According to Ref.~\cite{Tomalak:2020zfh}, hadronic contributions are the dominant source of the uncertainty in CEvNS cross sections at low (anti)neutrino beam energy and low nuclear recoils. At higher energies, nuclear modeling uncertainties become dominant. Higher-order Coulomb corrections in CEvNS are found to be negligible~\cite{Plestid:2023mta}. Notably, light-quark contributions cancel in electron-muon, electron-tau, and muon-tau flavor ratios of (anti)neutrino-induced neutral-current processes~\cite{Tomalak:2020zfh,Sehgal:1985iu,Botella:1986wy}, which makes these ratios particularly clean probes of new physics.

Hadronic contributions of interest enter through loop effects and are therefore suppressed by $\alpha$. In Refs.~\cite{Tomalak:2019ibg,Hill:2019xqk,Tomalak:2020zfh}, the light-quark contribution to neutral-current processes at momentum transfer $q$ is expressed in terms of charge-charge and charge-isospin current-current correlators $\mathrm{\Pi}_{\gamma \gamma} \left( q^2 \right)$ and $\mathrm{\Pi}_{3 \gamma} \left( q^2 \right)$, respectively,
\begin{align}
	\left( q^\mu q^\nu -q^2 g^{\mu \nu} \right) \mathrm{\Pi}_{\gamma \gamma} \left( q^2 \right) & = 4 i \pi^2 \int \mathrm{d}^d x e^{i q \cdot x} \langle 0 | T \left[ J_\gamma^\mu \left( x \right) J_\gamma^\nu \left( 0 \right) \right] | 0 \rangle, \label{eq:charge_charge_correlator} \\
	\left( q^\mu q^\nu -q^2 g^{\mu \nu} \right) \mathrm{\Pi}_{3 \gamma} \left( q^2 \right) & = 4 i \pi^2 \int \mathrm{d}^d x e^{i q \cdot x} \langle 0 | T \left[ J_3^\mu \left( x \right) J_\gamma^\nu \left( 0 \right) \right] | 0 \rangle, \label{eq:charge_isospin_correlator}
\end{align}
that are determined from the coordinate-space correlation functions of the electromagnetic and isospin quark currents $J_\gamma^\mu \left( x \right) = \overline{q} \gamma^\mu Q q \left( x \right)$ and $J_3^\mu \left( x \right) = \overline{q} \gamma^\mu \tau^3 q \left( x \right)$, respectively, where $q(x)$ denote the quark fields in flavor space, in $d$-dimensional space-time, and $\tau^3$ is the matrix representation for the third component of the isospin. For low-energy processes such as coherent elastic (anti)neutrino-nucleus scattering with decay-at-rest sources or reactor antineutrinos, and elastic (anti)neutrino-electron scattering below $\sim$1 TeV, it is sufficient to evaluate these correlators at zero momentum transfer: $\mathrm{\Pi}_{\gamma \gamma} \left( q^2 \right) \to \mathrm{\Pi}_{\gamma \gamma} \left( 0 \right)$ and $\mathrm{\Pi}_{3 \gamma} \left( q^2 \right) \to \mathrm{\Pi}_{3 \gamma} \left( 0 \right)$. Within the $\mathrm{SU}\left(3 \right)_f$ symmetry approximation~\cite{Jegerlehner:1985gq,Jegerlehner:2011mw}, Ref.~\cite{Tomalak:2019ibg} estimates the charge-isospin correlator at the scale $\mu = 2~\mathrm{GeV}$ for $n_f = 3$ quark flavors as $\hat{\Pi}^{\left( 3 \right)}_{3 \gamma} \left( 0 \right) |_{\mu = 2~\mathrm{GeV}} = \left( 1 \pm 0.2 \right)\hat{\Pi}^{\left( 3 \right)}_{\gamma \gamma} \left( 0 \right) |_{\mu = 2~\mathrm{GeV}}$, with $20\%$ uncertainty of the flavor symmetry ansatz and $\hat{\Pi}^{\left( 3 \right)}_{\gamma \gamma} \left( 0 \right) |_{\mu = 2~\mathrm{GeV}} = 3.597\pm0.021$~\cite{Erler:1998sy,Erler:2004in,Erler:2017knj,Proceedings:2019vxr} is obtained from a dispersive analysis of $e^{+} e^{-}$ cross-section data and measurements of hadronic $\tau$ decays, combined with a perturbative treatment of the high-energy contribution. The limited precision in $\hat{\Pi}^{(3)}_{3\gamma}(0)$ translates into a sizable uncertainty at the per-mille level in all low-energy neutral-current (anti)neutrino scattering processes.

At $q^2 = 0$, chiral perturbation theory (ChPT) provides a controlled EFT framework for evaluating hadronic contributions. In this work, we study the hadronic structure encoded in $\Pi_{3\gamma}$ within ChPT and demonstrate a significant reduction in the associated theoretical uncertainty.

Following the spurion technique employed in the meson and baryon sectors~\cite{Gasser:1983yg,Gasser:1984gg,Moussallam:1997xx,Scherer:2002tk,Descotes-Genon:2005wrq,HillerBlin:2016jpb,Cirigliano:2023fnz,Cirigliano:2024nfi}, we determine matrix elements of quark currents by taking functional derivatives of the generating functional $W$ with respect to the electromagnetic spurion field ${\bf q}_V$, excluding the explicit charge matrix $Q$, coupled to the background photon field $A^\mu$, and the isovector component of the external vector source $v^\mu$. This procedure yields the following:
\begin{align}
	\langle 0 | T \left[ J_\gamma^\mu \left( x \right) J_\gamma^\nu \left( 0 \right) \right] | 0 \rangle &= - \frac{4 i}{e^2} \langle 0 | T \left[ \frac{\delta^2 W \left( {\bf q}_V A^\lambda, v^\lambda \right)}{\delta \left( {\bf q}_V A_\mu \left(x \right) \right) \delta \left( {\bf q}_V A_\nu \left( 0 \right) \right)} \right]| 0 \rangle, \label{eq:charge_charge} \\
	\langle 0 | T \left[ J_3^\mu \left( x \right) J_\gamma^\nu \left( 0 \right) \right]| 0 \rangle &= \frac{4 i }{e} \langle 0 | T \left[ \frac{\delta^2 W \left( {\bf q}_V A^\lambda, v^\lambda \right)}{\delta v_\mu^3 \left( x \right) \delta \left( {\bf q}_V A_\nu \left( 0 \right) \right)} \right] | 0 \rangle \label{eq:isospin_charge}.
\end{align}

We evaluate the corresponding derivatives using the leading-order chiral Lagrangian in the meson sector as
\begin{align}
	\frac{\delta W \left( {\bf q}_V A^\lambda, v^\lambda \right)}{\delta \left( {\bf q}_V A_\mu \left( x \right) \right)} & = \frac{i F^2_0}{2} < \left( u^\dagger \partial^\mu u - u \partial^\mu u^\dagger \right) \left( u Q u^\dagger - u^\dagger Q u \right) >, \label{eq:charge_current} \\
	\frac{\delta W \left( {\bf q}_V A^\lambda, v^\lambda \right)}{\delta v^3_\mu \left( x \right)} & = - \frac{i F^2_0}{2} < \left( u^\dagger \partial^\mu u - u \partial^\mu u^\dagger \right) \left( u \frac{\tau^3}{2} u^\dagger - u^\dagger \frac{\tau^3}{2} u \right) >, \label{eq:isospin_current}
\end{align}
where $Q$ and $\tau^3$ denote the meson charge and isospin matrices, respectively. The chiral field is defined as $u = e^{\frac{i \Phi^a \tau^a}{2 F_0}}$, with meson fields $\Phi^a$ and generators $\tau^a$ of the fundamental representation of the Lie algebra of the chiral symmetry group. The constant $F_0 \approx 93~\mathrm{MeV}$ is the meson decay constant.

In $\mathrm{SU}\left( 2 \right)$ ChPT, the meson charge matrix is expressed in terms of the isospin matrix as $Q = \frac{1}{2} + \frac{\tau^3}{2}$. As a result, equations~(\ref{eq:charge_current}) and~(\ref{eq:isospin_current}) become identical, up to an overall sign. Adopting the bottom-up construction of ChPT, we thus impose an exact relation between $v^\mu_3 $ and $e {\bf q}_V A^\mu$ in all ChPT building blocks that preserves the connection between equations~(\ref{eq:charge_current}) and~(\ref{eq:isospin_current}) order-by-order. Consequently, $\mathrm{\Pi}_{3 \gamma} \left( q^2 \right) = \mathrm{\Pi}_{\gamma \gamma} \left( q^2 \right)$ at all orders in chiral and electromagnetic expansions, for sufficiently low $q^2$ where ChPT is valid. This helps us to avoid large hadronic uncertainties from $\mathrm{\Pi}_{3 \gamma} \left( q^2 \right)$ in low-energy neutral-current (anti)neutrino scattering. General EFT arguments allow us to perform $\mathrm{SU}\left( 2 \right)$ ChPT analysis without explicit evaluations. Recent lattice-QCD calculations for the momentum-transfer dependence of $\mathrm{\Pi}_{3 \gamma} \left( q^2 \right)$ in Refs.~\cite{Burger:2015lqa,Francis:2015grz,Ce:2018ziv,Ce:2022eix,Conigli:2025qvh} validate the equality $\mathrm{\Pi}_{3 \gamma} \left( q^2 \right) \approx \mathrm{\Pi}_{\gamma \gamma} \left( q^2 \right)$ within $5$-$10\%$ accuracy level.

In contrast, in $\mathrm{SU}\left( 3 \right)$ ChPT, the meson charge matrix becomes $Q = \frac{\lambda^3}{2} + \frac{\lambda^8}{2 \sqrt{3}}$. Here, the identity $\Pi_{3\gamma}(q^2) = \Pi_{\gamma\gamma}(q^2)$ is not exact, since the Gell-Mann matrix $\lambda^8$, a generator of the $\mathrm{U}\left(1 \right)$ hypercharge, satisfies $u \lambda^8 u^\dagger \neq u^\dagger \lambda^8 u$. We therefore perform an explicit leading-order calculation of both $\Pi_{\gamma\gamma}$ and $\Pi_{3\gamma}$, extending the analysis of Ref.~\cite{Czarnecki:1995fw}, and introduce the charge-charge $c_{Q Q}$ and charge-hypercharge $c_{QY}$ counterterms. The charge-hypercharge counterterm $c_{QY}$ is assumed to vanish at the renormalization scale $\mu_\chi \lesssim m_{K^{+}}$, where kaons decouple. The resulting ChPT expressions for the correlation functions $\mathrm{\Pi}_{\gamma \gamma}$ and $\mathrm{\Pi}_{3 \gamma}$ are given by
\begin{align}
	\mathrm{\Pi}_{\gamma \gamma} \left( q^2 \right) & = c_{QQ} \left( \mu_\chi \right) + \hat{B}_0 \left(\mu_\chi^2,~q^2,~m^2_{\pi^+} \right) + \hat{B}_0 \left(\mu_\chi^2,~q^2,~m^2_{K^+} \right) , \label{eq:charge_current_SU3} \\
	\mathrm{\Pi}_{3 \gamma} \left( q^2 \right) & = c_{Q Q} \left( \mu_\chi \right) +  \hat{B}_0 \left(\mu_\chi^2,~q^2,~m^2_{\pi^+} \right) + \hat{B}_0 \left(\mu_\chi^2,~q^2,~m^2_{K^+} \right) + c_{QY} \left( \mu_\chi \right) - \frac{1}{2} \hat{B}_0 \left(\mu_\chi^2,~q^2,~m^2_{K^+} \right), \label{eq:isospin_current_SU3}
\end{align}
where $\hat{B}_0$ denotes the bubble-type combination of the standard Passarino-Veltmann loop functions for degenerate internal masses~\cite{Passarino:1978jh,Ellis:2007qk}:
\begin{align}
	\hat{B}_0 \left(\mu_\chi^2,~q^2,~m^2 \right) &= \frac{\mu^{4-d}_\chi}{d-1} \int \frac{4 \mathrm{d}^d L}{i \left(2 \pi \right)^{d-2}} \frac{1 - \frac{4 m^2}{q^2}}{ \left( L^2 - m^2 \right) \left( \left( L + q \right)^2 - m^2 \right)} + \frac{d - 2}{q^2} \frac{\mu^{4-d}_\chi}{d-1} \int \frac{4 \mathrm{d}^d L}{i \left(2 \pi \right)^{d-2}} \frac{2}{L^2 - m^2}, \label{eq:bubble}
\end{align}
with the renormalized value in the $\overline{\mathrm{MS}}_\chi$ scheme:
\begin{align}
	\hat{B}_0 \left(\mu_\chi^2,~q^2,~m^2 \right) &= \frac{1}{3} \left( 1 - \ln \frac{\mu^2_\chi}{m^2} \right) - \frac{2}{9} -  \frac{2 \left( 1 - \frac{4 m^2}{q^2} \right)}{3} + \frac{\left( 1 - \frac{4 m^2}{q^2} \right)^{\frac{3}{2}}}{3} \ln \frac{\sqrt{ 1 - \frac{4 m^2}{q^2}} + 1}{\sqrt{ 1 - \frac{4 m^2}{q^2}} - 1}. \label{eq:bubble_renormalized}
\end{align}

In $\mathrm{SU}\left( 2 \right)$ ChPT, expressions~(\ref{eq:charge_current_SU3}) and~(\ref{eq:isospin_current_SU3}) have different values and scale-dependence for the charge-charge counterterm $c_{QQ} \left( \mu_\chi \right)$, the counterterm $c_{QY} \left( \mu_\chi \right)$ and kaon contributions are absent. We determine $c^{\mathrm{SU(3)}}_{QQ} \left( \mu_\chi \right)$ and $c^{\mathrm{SU(3)}}_{QY} \left( \mu_\chi \right)$ in terms of $c^{\mathrm{SU(2)}}_{QQ} \left( \mu_\chi \right)$ by matching the $\mathrm{SU}\left( 3 \right)$ ChPT expressions~(\ref{eq:charge_current_SU3}) and~(\ref{eq:isospin_current_SU3}) to the $\mathrm{SU}\left( 2 \right)$ ChPT results at the kinematical point $q^2 = 0$ when both theories should give identical amplitudes. Technically, we integrate out kaon fields at the scale around the kaon mass and obtain
\begin{align}
	c^{\mathrm{SU(3)}}_{QQ} \left( \mu_\chi \right) &= c^{\mathrm{SU(2)}}_{QQ} \left( \mu_\chi \right) - \hspace{-0.03cm} \hat{B}_0 \left(\mu_\chi^2,~0,~m^2_{K^+} \right) \left[ 1 + \mathcal{O} \left( \frac{m^2_{K^+}}{\left( 4 \pi F_0 \right)^2}\right) \right], \label{eq:charge_charge_counterterm_SU3} \\
	c^{\mathrm{SU(3)}}_{QY} \left( \mu_\chi \right) &= \frac{1}{2} \hat{B}_0 \left(\mu_\chi^2,~0,~m^2_{K^+} \right) \left[ 1 + \mathcal{O} \left( \frac{m^2_{K^+}}{\left( 4 \pi F_0 \right)^2}\right) \right].\label{eq:charge_isospin_counterterm_SU3}
\end{align}
The counterterms $c_{QQ}$ and $c_{QY}$ at leading order in ChPT can be also expressed in terms of the ChPT low-energy coupling constants~\cite{Gasser:1983yg,Gasser:1984gg,Scherer:2002tk} as
\begin{align}
c_{QQ}^{\mathrm{SU(3)}} \left( \mu_\chi \right) &= - 4 \left(4 \pi \right)^2 \left( L_{10} \left( \mu_\chi \right) + 2 H_1 \left( \mu_\chi \right) \right) < Q^2 >^{\mathrm{SU(3)}} = - \frac{8}{3} \left(4 \pi \right)^2 \left( L_{10} \left( \mu_\chi \right) + 2 H_1 \left( \mu_\chi \right) \right), \label{eq:charge_charge_counterterm_SU3_LECs} \\
c_{QY}^{\mathrm{SU(3)}} \left( \mu_\chi \right) &=  \frac{< Q \frac{\lambda_3}{2} >^{\mathrm{SU(3)}} - < Q^2 >^{\mathrm{SU(3)}}}{< Q^2 >^{\mathrm{SU(3)}}} c_{QQ}^{\mathrm{SU(3)}} \left( \mu_\chi \right) = \frac{2}{3} \left(4 \pi \right)^2 \left( L_{10} \left( \mu_\chi \right) + 2 H_1 \left( \mu_\chi \right) \right), \label{eq:charge_isospin_counterterm_SU3_LECs} \\
c_{QQ}^{\mathrm{SU(2)}} \left( \mu_\chi \right) &= 16  \left(4 \pi \right)^2 h_2 \left( \mu_\chi \right)  < Q^2 >^{\mathrm{SU(2)}} = 16  \left(4 \pi \right)^2 h_2 \left( \mu_\chi \right), \label{eq:charge_charge_counterterm_SU2_LECs}
\end{align}
with results for the charge-charge correlator and a few individual isospin components available in Refs.~\cite{Gasser:1984gg,Knecht:1994ug,Golowich:1995kd,Maltman:1995jg,Moussallam:1997xx,Maltman:1997tr,Amoros:1999dp,Durr:1999dp,Gasser:2007sg,Pich:2008jm,Gonzalez-Alonso:2008sva,Gonzalez-Alonso:2008oox,Gasser:2009hr,Gasser:2010zz,Bijnens:2011xt,Golterman:2014nua,Boyle:2014pja,Golterman:2017ljr,Golterman:2020pyx,Lellouch:2025rnz}. Equations~(\ref{eq:charge_isospin_counterterm_SU3}) and~(\ref{eq:charge_isospin_counterterm_SU3_LECs}) are in agreement within uncertainty region for $L_{10}$ and $H_1$~\cite{Gonzalez-Alonso:2008sva,Boito:2015fra}.
\begin{figure}[htb]
	\centering
	\includegraphics[width=0.49\textwidth]{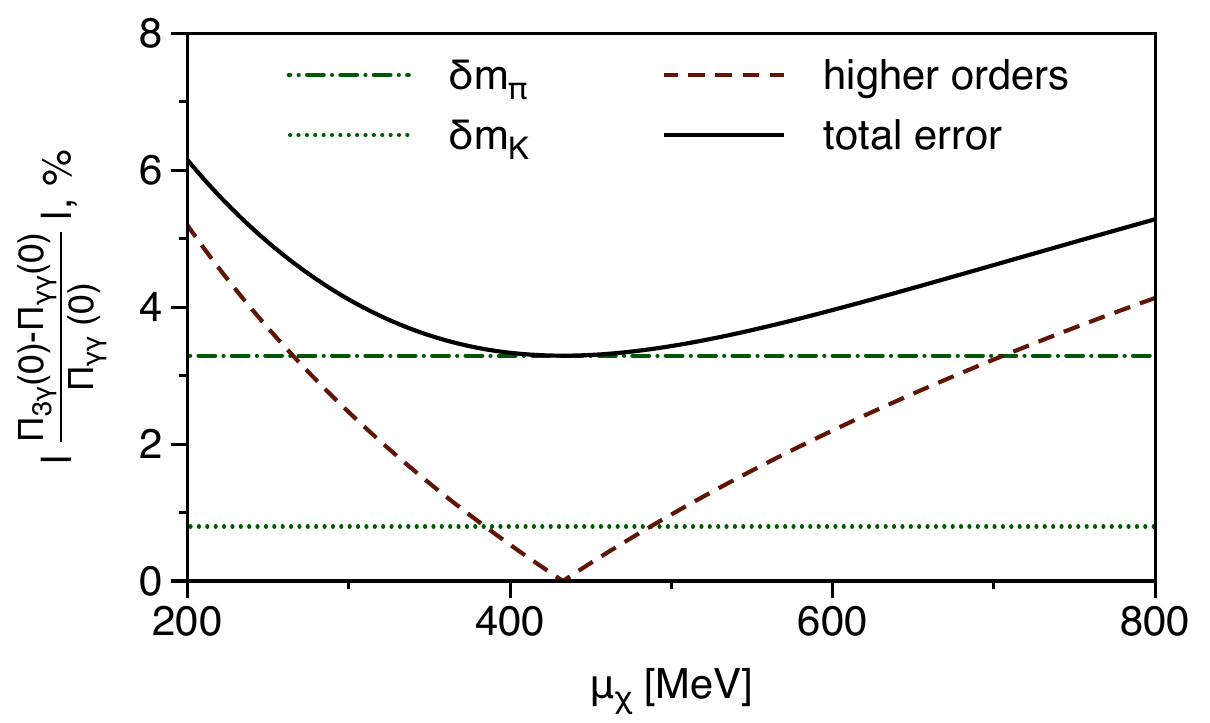}
	\caption{Error sources in the relative difference between charge-charge and charge-isospin correlation functions, $\frac{\mathrm{\Pi}_{3 \gamma} \left( 0 \right) - \mathrm{\Pi}_{\gamma \gamma} \left( 0 \right)}{\mathrm{\Pi}_{\gamma \gamma} \left( 0 \right)}$, at one loop in $\mathrm{SU}\left( 3 \right)$ ChPT analysis are shown as functions of the renormalization scale $\mu_\chi$. The relative error from the pion(kaon) mass splitting is shown as a green dash-dotted(dotted) line, respectively. The higher-order error is represented by a red dashed line. The total uncertainty, shown as a black solid line, is computed by summing the errors in quadrature. It is automatically reduced to zero in $\mathrm{SU}\left( 2 \right)$ ChPT analysis.} \label{fig:SU3_error}
\end{figure}

We consider two primary sources of uncertainty in our leading-order analysis: (i) isospin-violating mass differences arising from quantum electrodynamics, quantified by the ratio $\frac{m_{\pi^+} - m_{\pi^0}}{m_{\pi^+} }$; and (ii) higher-order chiral corrections, which we estimate by multiplying the leading-order result by the chiral expansion parameter $\frac{m^2_K}{\left( 4 \pi F_0 \right)^2}$. We combine these uncertainties in quadrature to estimate the total error. Fig.~\ref{fig:SU3_error} illustrates the resulting uncertainty in the ratio $\frac{\mathrm{\Pi}_{3 \gamma} \left( 0 \right) - \mathrm{\Pi}_{\gamma \gamma} \left( 0 \right)}{\mathrm{\Pi}_{\gamma \gamma} \left( 0 \right)}$ as a function of the chiral scale $\mu_\chi$, using numerical inputs from Ref.~\cite{ParticleDataGroup:2020ssz}. At leading order, the Dashen’s theorem~\cite{Dashen:1969eg} predicts equal electromagnetic contributions to the mass splittings of pion and kaon. The pion mass splitting is dominated by electromagnetic effects~\cite{Socolow:1965zz,Das:1967it,Ecker:1988te,Bardeen:1988zw,Duncan:1996sq,Knecht:1997jw}, with only a small strong interaction component, making it suitable for error estimates. In contrast, the kaon mass splitting, $\frac{m_{K^+} - m_{K^0}}{m_{K^+}} \ll \frac{m_{\pi^+} - m_{\pi^0}}{m_{\pi^+} }$, also shown in Fig.~\ref{fig:SU3_error}, receives significant contributions from QCD, indicating a strong violation of the Dashen’s theorem~\cite{Langacker:1973udm,Maltman:1990mq,Donoghue:1993hj,Bijnens:1993ae,Neufeld:1994eg,Urech:1994hd,Neufeld:1995mu,Baur:1995ig,Donoghue:1996zn,Gao:1996sa,Bijnens:1996kk,Moussallam:1997xx,Moussallam:1998za,Gao:1999ww,Bijnens:2006mk,Blum:2007cy,Hayakawa:2008an,MILC:2009ltw,Blum:2010ym,Portelli:2011trb,deDivitiis:2013xla,Portelli:2015wna,Horsley:2015eaa,MILC:2015vfd,Fodor:2016bgu,Giusti:2017dmp,MILC:2018ddw,Rowe:2023jlt,FlavourLatticeAveragingGroupFLAG:2024oxs}. We expect the cancellation of higher-order uncertainty as well as isospin-violating contributions when integrating out kaon fields and determining the counterterm $c_{QY} \left( \mu_\chi \right)$ order-by-order in chiral expansion in the relation between hadronic correlators $\hat{\Pi}^{\left( 3 \right)}_{3 \gamma} \left( 0 \right) |_{\mu = 2~\mathrm{GeV}}$ and $\hat{\Pi}^{\left( 3 \right)}_{\gamma \gamma} \left( 0 \right) |_{\mu = 2~\mathrm{GeV}}$.

After refining the most uncertain hadronic contributions, we update the cross-section uncertainties for elastic (anti)neutrino–electron and coherent elastic (anti)neutrino–nucleus scattering, as presented in Tables~\ref{tab:errors_neutrino_electron} and~\ref{tab:errors_CEvNS_Argon}, based on Refs.~\cite{Tomalak:2019ibg,Tomalak:2020zfh}. These updates substantially reduce the theoretical uncertainty in elastic (anti)neutrino–electron scattering across all energies and in coherent elastic (anti)neutrino–nucleus scattering at low (anti)neutrino beam energies. Compared to earlier estimates~\cite{Tomalak:2019ibg,Tomalak:2020zfh}, this paper reduces hadronic errors by a factor $\sim35$.
\begin{table}
	\centering
	\caption{\textbf{Error budget in elastic (anti)neutrino-electron scattering.}
	$1\sigma$ contributions to the relative error (in $\%$) of the total (anti)neutrino-electron scattering cross section for an incident (anti)neutrino energy much larger than the electron mass $m_e$, i.e., $E_\nu \gg m_e$.}
	\label{tab:errors_neutrino_electron}
	\begin{tabular}{|l|c|c|c|c|c|c|c|}
	\hline          
	& Hadronic & Quark & Pert. & Total~\cite{Tomalak:2019ibg} & Total \\
	\hline
	$\nu_\mu e \to \nu_\mu e(\gamma)$  &0.011 &0.068 & $\lesssim$ 0.008 & 0.37 & 0.069\\
	$\bar{\nu}_\mu e \to \bar{\nu}_\mu e(\gamma)$  &0.009 & 0.112 & $\lesssim$ 0.005 & 0.33 & 0.112 \\
	$\nu_e e \to \nu_e e(\gamma)$  &0.008 & 0.028 & $\lesssim$ 0.007 & 0.26 & 0.030 \\
	$\bar{\nu}_e e \to \bar{\nu}_e e(\gamma)$  &0.011 & 0.044 & $\lesssim$ 0.007 & 0.36 & 0.045 \\
	\hline
	\end{tabular}
\end{table}

\begin{table}
	\centering
	\caption{\textbf{Error budget in coherent elastic neutrino-nucleus scattering.}
	$1\sigma$ contributions to the relative error (in $\%$) of the total CE$\nu_\mu$NS cross section on$~^{40} \mathrm{Ar}$ target for an incident $\nu_\mu$ neutrino energy $E_\nu$.}
	\label{tab:errors_CEvNS_Argon}
	\begin{tabular}{|c|c|c|c|c|c|c|c|}   
	\hline          
	$E_\nu,~\mathrm{MeV}$  &  Nuclear & Nucleon  & Hadronic & Quark  & Pert. & Total~\cite{Tomalak:2020zfh} & Total \\
	\hline
	50 & 4. & 0.06 & 0.016 & 0.13 & 0.08 & 4.05 & 4.01 \\
	30 &1.5 & 0.014 & 0.016 & 0.13 & 0.03 & 1.65 & 1.55 \\
	10 & 0.04 & 0.001 & 0.016 & 0.13 & 0.004 & 0.58 & 0.15 \\
	\hline
	\end{tabular}
\end{table}

In this work, we analyzed the hadronic contributions to charge-charge and charge-isospin correlation functions of quark currents using chiral perturbation theory. At low energies, we demonstrated that these correlators are identical within $\mathrm{SU}\left( 2 \right)$ ChPT, thereby reducing the corresponding hadronic uncertainty in low-energy neutral-current (anti)neutrino scattering processes by an order of magnitude without shifting central values previously obtained in the $\mathrm{SU}(3)_f$ approximation. The leading theoretical uncertainty in coherent elastic neutrino-nucleus scattering at very low energies and in elastic (anti)neutrino-electron scattering now arises from the unknown $\mathcal{O}(\alpha \alpha_s)$ matching contribution to the Wilson coefficients of neutral-current (anti)neutrino-quark four-fermion interactions. This error is at the level $\lesssim 1$ permille. Beyond its immediate phenomenological applications, our result has implications for parity-violating electron scattering and atomic parity violation and provides a benchmark for future lattice-QCD calculations of current-current correlation functions.

\section*{Acknowledgments}
I thank Richard J. Hill, Ryan Plestid, Vishvas Pandey, Pedro Machado, Emanuele Mereghetti, Wouter Dekens, Vincenzo Cirigliano for useful discussions while working on the related subjects, Jens Erler for useful communication, and Feng-Kun Guo for their invaluable feedback on this manuscript. \textsc{FeynCalc}~\cite{Mertig:1990an,Shtabovenko:2016sxi}, \textsc{LoopTools}~\cite{Hahn:1998yk}, \textsc{Wolfram Mathematica}~\cite{Mathematica}, and \textsc{DataGraph}~\cite{JSSv047s02} were extremely useful in this work. \textsc{Wolfram Mathematica} and \textsc{DataGraph} Licenses were sponsored by Los Alamos National Laboratory. This work is supported by the National Science Foundation of China under Grants No. 12347105, No. 12447101.

\bibliography{charge_isospin}{}

\end{document}